\documentclass[10.5pt,letterpaper,fleqn, twocolumn]{article}

\usepackage{graphicx}
\usepackage{makeidx}
\usepackage{multicol}
\usepackage{crop}
\usepackage{amstext,amsthm}
\usepackage{amsmath,amssymb}
\usepackage{bm}
\usepackage{natbib}

\usepackage{multicol}
\usepackage{pslatex}
\usepackage[strict]{changepage}
\newcommand{\be}{\begin{equation}}
\newcommand{\ee}{\end{equation}}
\newcommand{\bea}{\begin{eqnarray}}
\newcommand{\eea}{\end{eqnarray}}
\newcommand{\bne}{\begin{equation*}}
\newcommand{\ene}{\end{equation*}}
\newcommand{\bi}{\begin{itemize}}
\newcommand{\ei}{\end{itemize}}

\newcommand{\bbm}{\begin{bmatrix}}
\newcommand{\ebm}{\end{bmatrix}}

\newcommand{\mr}{\mathrm}

\usepackage[top=0.75in, bottom=0.5in, left=0.83in, right=0.83in]{geometry}
\usepackage[hang,splitrule]{footmisc}





\usepackage{titlesec}		
\usepackage{subcaption}		
\usepackage{indentfirst}	

\newcommand{\eps}{\epsilon}

\newenvironment{myabstract}
{
	\vspace{0.2in}
	\parindent=0cm \textit{Abstract}
	\parindent=0.5cm \hangindent=0.5cm \linebreak
}

\newenvironment{mykeywords}
{
	\vspace{0.2in}
	\parindent=0cm \textit{Keywords}
	
	\parindent=0.5cm 
}

\titleformat{\section}
{\normalfont\bfseries}
{}{0.0pt}{\parindent=0cm}[\parindent=0.5cm]

\titlespacing*{\section}
{0pt}{1.5\baselineskip}{0.5\baselineskip}

\titleformat{\subsection}
{\normalfont\itshape}
{}{0.0pt}{\parindent=0cm}[\parindent=0.5cm]

\titlespacing*{\subsection}
{0pt}{1.0\baselineskip}{0.5\baselineskip}

\begin{document}


\twocolumn[{%
	
\phantom\\
\vspace{0.5in}
\begin{center}
\Large{\textbf{MODEL PREDICTIVE CONTROL TUNING BY MONTE CARLO SIMULATION AND CONTROLLER MATCHING}}\\
\end{center}
\vspace{0.2in}

\begin{center}
Morten Ryberg Wahlgreen $^{\mr{a}}$, John Bagterp J{\o}rgensen $^{\mr{a,}}$\footnotemark and Mario Zanon $^{\mr{b}}$ \\

\vspace{0.10in}

$^{\mr{a}}$ Department of Applied Mathematics and Computer Science, Technical University of Denmark, DK-2800 Kgs. Lyngby, Denmark

\vspace{0.10in}
$^{\mr{b}}$ IMT School for Advanced Studies Lucca, IT-55100 Lucca, Italy
\end{center}

\begin{myabstract}
	This paper presents a systematic method for the selection of the Model Predictive Control (MPC) stage cost. We match the MPC feedback law to a proportional-integral (PI) controller, which we efficiently tune by high-performance Monte Carlo (MC) simulation. The PI tuning offers a wide range of tuning possibilities that is then inherited by the MPC design. The MC simulation tuning of the PI controller is based on the minimization of two different objectives; 1) the 2-norm tracking error, and 2) a bi-objective consisting of the 2-norm tracking error and a 2-norm input rate of movement penalty. We apply the method to design MPC for an exothermic chemical reaction conducted in an adiabatic continuous stirred tank reactor (CSTR). The process is of interest as the nonlinear dynamics result in a desired operating point very close to a constraint. Our MPC design includes stage costs automatically designed to match the tuned PI controllers, hard input constraints, and a soft output constraint. Stochastic simulation results show that both the PI controller and the MPC can track the desired operating point. However, the MPC shows reduced output constraint violation compared to the PI controller. As such, the MPC design method successfully combines the efficient tuning of the PI controller with the constraint handling properties of MPC.







\end{myabstract}

\begin{mykeywords}
	Model Predictive Control, Controller Matching, Automatic Monte Carlo Simulation Tuning.
\end{mykeywords}

\vspace{0.2in}
}]

\footnotetext[1]{\parindent=0cm \small{Corresponding author: J. B. J{\o}rgensen (E-mail: {\tt\small jbjo@dtu.dk}).}}

%
%


\section{Introduction} 
\label{sec:introduction}

Model Predictive Control (MPC) is an advanced control technology, that is widely applied in the industry \citep{qin:2003}. MPC offers direct methods to handle constraints in the specific system. However, the lack of systematic tuning methods can make the selection of the MPC parameters cumbersome. On the contrary, simple linear controllers, such as proportional–integral (PI) controllers, offer systematic tuning options. Therefore, it has previously been proposed to match the MPC stage cost to tuned linear controllers \citep{diCairano:2009a, diCairano:2010a, zanon:2021a}. As such, one benefits from the systematic tuning of linear controllers in combination with the MPC ability to handle constraints. 

There exist a variety of tuning methods for linear controllers, see, e.g., \citep{bansal:2012} for proportional–integral–derivative (PID) controllers. Also, we have previously proposed an automatic tuning method for PID controllers based on high-performance Monte Carlo (MC) simulations of stochastic closed-loop systems \citep{wahlgreen:2021a}. This tuning method offers a systematic approach, where one can select any tuning objective, e.g., target tracking, input usages, etc. The method is even applicable to advanced controllers like MPC. However, the tuning process requires a large number of closed-loop simulations, which can be computationally expensive for MPC. The computational cost of MC simulations for direct MPC tuning will be significantly higher compared to MC simulation for tuning linear controllers.



In this paper, we propose an MPC design method based on matching the MPC feedback to a PI controller tuned by MC simulations. As such, the method offers a systematic and efficient approach to design the MPC stage cost. We demonstrate the design method on a simulation case study, where we consider an exothermic chemical reaction conducted in an adiabatic continuous stirred tank reactor (CSTR) \citep{wahlgreen:2020a, jorgensen:2020a}. We apply a nonlinear stochastic differential equation (SDE) model for the CSTR dynamics, which results in an operating point close to an output constraint. We track the optimal operating point with a PI controller and an MPC controller, where the PI controller is tuned based on MC simulation and the MPC is matched to the tuned PI controller. The MPC formulation includes hard input constraints and a soft output constraint to avoid infeasible Optimal Control Problems (OCPs) in the MPC. Our results show that MPC reduces output constraint violation compared to the PI controller while maintaining the tuned PI performance when output constraints are inactive. 

The remaining parts of the paper are organized as follows. First, we introduce the simulation model for the CSTR. Next, we present the discrete PI controller with anti-windup mechanism and the hard input and soft output constrained MPC formulation for stabilization of the CSTR. Then, we present the controller matching problem, show how to represent the PI controller as a linear controller for the system, and provide a short description of the MC simulation tuning. Finally, we present the tuning and simulation results and end with our conclusions.


\section{Simulation Model for the Adiabatic CSTR}
\label{sec:CSTR}

We consider an exothermic reaction conducted in an adiabatic CSTR \citep{wahlgreen:2020a, jorgensen:2020a}. 

\subsection{General ODE model for the CSTR}
A general ordinary differential equation (ODE) model for a non-constant volume CSTR is \citep{wahlgreen:2022a}
\begin{subequations} \label{eq:CSTR_general}
	\begin{align} 
		\frac{dV}{dt} &= e^{\top}F_{\mathrm{in}} - F_{\mathrm{out}}, \\
		\frac{dn}{dt} &= C_{\mathrm{in}}F_{\mathrm{in}} - c F_{\mathrm{out}} + RV,
	\end{align}
\end{subequations}
where $n$ is a vector of mole numbers, $c=n/V$ is a vector of concentrations, $C_{\mathrm{in}}$ is a matrix of inlet concentrations, $R$ is the production rate, $V$ is the volume of the CSTR, $F_{\mathrm{in}}$ is a vector of inlet stream flow rates, $F_{\mathrm{out}}$ is a scalar with the outlet stream flow rate, and $e$ is a vector of ones of proper dimension. The production rate is given as,
\begin{align}
	R = S^{\top}r,
\end{align}
where $S$ is the stoichiometric matrix and $r = r(c)$ is the reaction rate.

\subsection{Exothermic reaction conducted in a constant volume CSTR}
The medium in the constant volume CSTR consists of two components, $A$ and $B$, and have temperature, $T$, which we treat as an additional chemical component. As such, the vector $n$ consists of mole numbers (for $A$ and $B$) and the total internal energy (for $T$). The vector $c$ consists of the concentrations of $A$ and $B$, and the temperature $T = c_T$. The CSTR has constant volume and one inlet stream, i.e., $e^{\top} F_{\mathrm{in}} = F_{\mathrm{in}} = F_{\mathrm{out}} = F$. The stoichiometric matrix is, 
\begin{align}
	S = \begin{bmatrix}
		-1.0 & -2.0 & \beta
	\end{bmatrix},
\end{align}
where $\beta = -\Delta H_r/(\rho c_P)$, $\Delta H_r$ is the enthalpy of reaction, $\rho$ is the density of the mixture, and $c_P$ is the specific heat capacity. The rate of reaction is,
\begin{align}
	r(c) = k(c_T) c_A c_B,
\end{align}
with
\begin{align} \label{eq:Arrhenius}
	k(c_T) = k_0 \exp\left( -\frac{E_a}{R}\frac{1}{c_T} \right),
\end{align}
and $E_a/R$ denoting the activation energy. The inlet stream has the concentrations
\begin{align} \label{eq:inletConcentrations}
	C_{\mathrm{in}} = \begin{bmatrix}
		c_{A,\mathrm{in}} \\
		c_{B,\mathrm{in}} \\
		c_{T,\mathrm{in}} 
	\end{bmatrix} = \begin{bmatrix}
		1.6/2 \\
		2.4/2 \\
		273.65
	\end{bmatrix}.
\end{align}
Together, Eq. (\ref{eq:CSTR_general})-(\ref{eq:inletConcentrations}) forms a three-state ODE model, where the states are $n_A$, $n_B$, and $n_T$. We refer to \citep{wahlgreen:2020a} for the system parameters.

\subsection{One-state model}
At steady-state, the three-state model is exactly represented by a one-state model, where $c_A(c_T)$ and $c_B(c_T)$ are functions of the temperature given as \citep{wahlgreen:2020a},
\begin{subequations}
	\begin{align}
		c_A(c_T) &= c_{A,\mathrm{in}} + \frac{1}{\beta}( c_{T,\mathrm{in}} - c_{T} ), \\
		c_A(c_T) &= c_{B,\mathrm{in}} + \frac{2}{\beta}( c_{T,\mathrm{in}} - c_{T} ).
	\end{align}
\end{subequations}
The resulting inlet matrix and stoichiometric matrix are, 
\begin{align}
	C_{in} = \begin{bmatrix}
		c_{T,\mathrm{in}}
	\end{bmatrix}, \quad S = \begin{bmatrix}
	\beta
\end{bmatrix},
\end{align}
and the state is $n_T$.

\subsection{Stochastic differential equations}
We extend the general ODE formulation, Eq. (\ref{eq:CSTR_general}), to an SDE formulation with a stochastic diffusion term and disregard the volume equation since the volume is constant. The SDE is,
\begin{align}
	dn(t) = \left( C_{\mathrm{in}}F - c F + RV \right) dt + F \bar \sigma d\omega(t),
\end{align}
where $d\omega(t) \sim N_{idd}(0, I dt)$ is a standard Wiener process and the diffusion function, $F\bar\sigma$ with $\bar\sigma = \text{diag}([ 0; 0; \sigma_T ])$, models inlet temperature variations \citep{wahlgreen:2020a}.

\subsection{Stochastic continuous-discrete system}

Let $x = n$ be the states, $u = F$ be the inputs, $y_k = y(t_k)$ be discrete measurements corrupted by noise, $z$ be the output, $p$ be the parameters, and $v_k = v(t_k) \sim N_{idd}(0, R_v)$ be measurement noise. Then, we formulate the system as a stochastic continuous-discrete system,
\begin{subequations} \label{eq:SDE}
	\begin{alignat}{3}
		\begin{split}
			d x(t) &= f(t, x(t), u(t), p) dt 
			\\ & \qquad + \sigma(t, x(t), u(t), p) d\omega(t), \label{eq:CDS_states}
		\end{split} \\
		y(t_k) &= g(t_k,x(t_k)) + v(t_k), \\
		z(t) &= h(t, x(t)),
	\end{alignat}
\end{subequations}
where
\begin{subequations}
	\begin{align}
		f(t, x(t), u(t), p) &= C_{\mathrm{in}}F - c F + RV, \\
		\sigma(t, x(t), u(t), p) &= F \bar\sigma, \\
		g(t_k, x(t_k)) &= n_T/V = c_T, \label{eq:g}  \\
		h(t, x(t) ) &= n_T/V = c_T. \label{eq:h}
	\end{align}
\end{subequations}
We assume that discrete measurements are available with sampling time, $T_s$.

\subsection{Operation of the CSTR}

\begin{figure}[tb]
	\centering
	\includegraphics[width=0.45\textwidth]{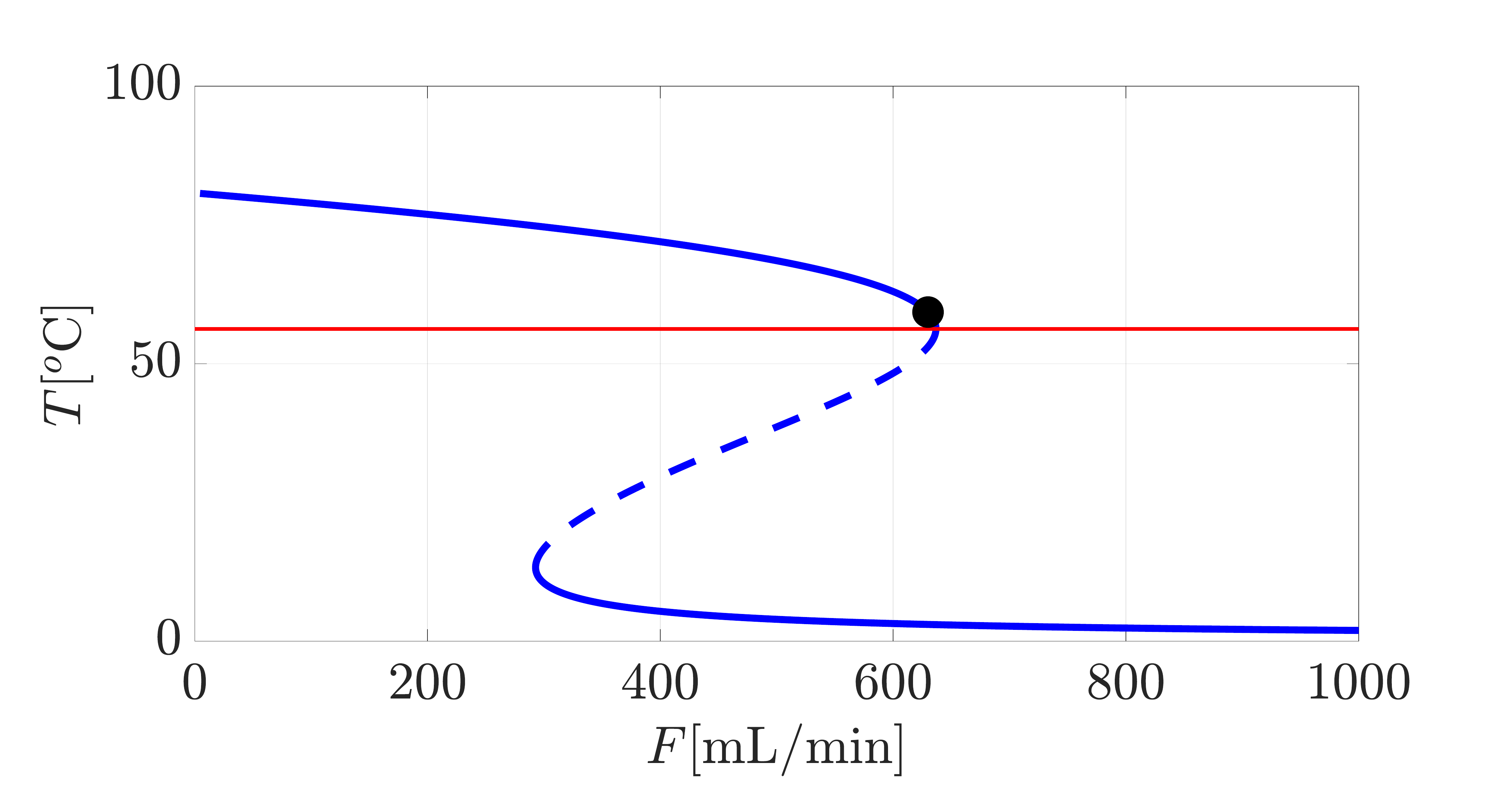}
	\caption{Steady state plot for the CSTR. Black dot: desired operation point. Red line: lower temperature constraint. }
	\label{fig:steadyState}
\end{figure}

Figure \ref{fig:steadyState} presents the steady-states of the CSTR within the flow rate limits, $F_{\min} = 0$ [mL/min] and $F_{\max} = 1000$ [mL/min] \citep{wahlgreen:2020a}. The desired operating point has the temperature steady-state $\bar T = 59.30$ [$^{\text{o}}$C] achieved at a flow rate of $\bar F = 630$ [mL/min]. The lower temperature constraint is $T_{\text{min}} = 57.26$ [$^{\text{o}}$C].


\section{PI Controller}
\label{sec:PI}

We stabilize the CSTR at the reference temperature, $\bar T$, with a PI controller. The discrete PI controller with anti-windup mechanism is given as,
\begin{subequations} \label{eq:PI}
	\begin{align}
		e_k &= \bar y_k - y_k, \\
		P_k &= k_P e_k, \\
		I_k &= \hat I_{k-1} + T_s k_I e_k, \\
		\hat u_k &= \bar u + P + I, \\
		u_k &= \max( u_{\text{min}}, \min( u_{\max}, \hat u_k ) ), \label{eq:PI_clip} \\
		I_{aw,k} &= T_s k_{aw}( u_k - \hat u_k ), \label{eq:PI_aw} \\
		\hat I_{k} &= I_k + I_{aw, k} \label{eq:PI_Ihat}.
	\end{align}
\end{subequations}
The anti-windup mechanism, Eq. (\ref{eq:PI_aw}), ensures reasonable integrator behavior, when the PI response saturates the input, $u_k$, at $u_{\min}$ or $u_{\max}$.


\section{Model Predictive Controller}
\label{sec:mpc}

We also stabilize the CSTR with a linear MPC (LMPC). At time $t_j$, the LMPC solves the hard input and soft output constrained OCP,
\begin{subequations} \label{eq:mpc}
	\begin{align}
		\min_{u, x} \quad & \phi_j(u,x) + \phi_{\eps, j}(u,x), \\
		s.t. \quad & x_j = \hat x_{j|j} \\
		& x_{j+k+1} = Ax_{j+k} + Bu_{j+k},  \\
		& z_{j+k} = C_z x_{j+k}, \\
		& u_{\min, j+k} \leq u_{j+k} \leq u_{\max, j+k}, \\
		& z_{j+k} \geq z_{\min, j+k} - \eps_{l, j+k}, \\
		& z_{j+k} \leq z_{\max, j+k} + \eps_{u, j+k},
	\end{align}
\end{subequations}
where $\hat x_{j|j}$ is an estimate of the states $x_j$ and

\begin{subequations} 
	\begin{alignat}{3}
		\begin{split}
			\phi_j(u,x) &= \sum_{k = 0}^{N-1} \begin{bmatrix}
				x_{j+k} \\
				u_{j+k}
			\end{bmatrix}^{\top} \begin{bmatrix}
				Q & S^{\top} \\
				S & R
			\end{bmatrix} \begin{bmatrix}
				x_{j+k} \\
				u_{j+k}
			\end{bmatrix} \\
		& \qquad + x_{j+N}^{\top} P x_{j+N}, 
		\end{split} \label{eq:stageCost} \\
		\begin{split}
			\phi_{\eps,j}(u,x) &= \sum_{k = 0}^N \left( \begin{bmatrix}
				\eps_{l, j+k}	\\
				\eps_{u, j+k}
			\end{bmatrix}^{\top} \begin{bmatrix}
				Q_{\eps_l} \\
				& Q_{\eps_u}
			\end{bmatrix}\begin{bmatrix}
				\eps_{l, j+k}	\\
				\eps_{u, j+k}
			\end{bmatrix} \right. \\
			& \left. \qquad + \begin{bmatrix}
				q_{\eps_l} \\
				q_{\eps_u}
			\end{bmatrix}^{\top} \begin{bmatrix}
				\eps_{l, j+k}	\\
				\eps_{u, j+k}
			\end{bmatrix} \right).
		\end{split}
	\end{alignat}
\end{subequations}

\section{ MPC Stage Cost Design }
\label{sec:cm}

We design the MPC stage cost, Eq. (\ref{eq:stageCost}), by controller matching to a well-tuned stabilizing PI controller.



\subsection{Controller matching problem}
The controller matching problem is formulated as the semi-definite programming (SDP) \citep{zanon:2021a},
\begin{subequations} \label{eq:SDP}
	\begin{align}
		\min_{\Gamma, P, \beta} \quad &\beta, \\
		s.t. \quad & \beta I \succeq H_{\Gamma} + H_P \succeq I,
	\end{align}
\end{subequations}
where 
\begin{subequations}
	\begin{align}
		H_{\Gamma} &= \begin{bmatrix}
			\hat K^{\top} \Gamma \hat K^{\top} & \hat K^{\top} \Gamma \\
			\Gamma \hat K & \Gamma
		\end{bmatrix}, \\
		H_{P} &= -\begin{bmatrix}
			A^{\top} P A - P & A^{\top} P B \\
			B^{\top} P A & B^{\top} P B
		\end{bmatrix},
	\end{align}
\end{subequations}
and $\hat K$ is the stabilizing linear controller feedback matrix to be matched such that,
\begin{align} \label{eq:linearController}
	u &= - \hat K x.
\end{align}
The OCP, Eq. (\ref{eq:mpc}), with stage cost matrices $Q$, $R$, and $S$,
\begin{subequations}
	\begin{align}
		Q &= \hat K^{\top} \Gamma \hat K + P - A^{\top} P A, \\
		R &= \Gamma - B^{\top} P B, \\
		S &= \Gamma \hat K - B^{\top} P A,
	\end{align}
\end{subequations}
produces the same response as the linear controller, Eq. (\ref{eq:linearController}), when inequality constrains are inactive.

\subsection{Linearization of the one-state model}
The matching problem, Eq. (\ref{eq:SDP}), requires a model in linear discrete state-space form,
\begin{subequations} \label{eq:linearStateSpace}
	\begin{align} 
		x_{k+1}^l &= Ax_k^l + Bu_k^l, \\
		y_k^l &= Cx_k^l.
	\end{align}
\end{subequations}
We apply linearization of the continuous model at the operation steady state, $(x_s, u_s)$, to obtain the continuous state-space matrices,
\begin{align}
	A^c &= \frac{\partial f}{\partial x}(x_s, u_s), & B^c = \frac{\partial f}{\partial u}(x_s, u_s), 
\end{align}
and exact discretization to obtain the discrete state-space matrices,
\begin{align}
	\begin{bmatrix}
		A & B \\
		0 & I
	\end{bmatrix} = \text{exp}\left( \begin{bmatrix}
		A^c & B^c \\
		0 	& 0 
	\end{bmatrix} T_s \right).
\end{align}
We base the MPC on the one-state model, as it is exact at steady-state. Linearization and discretization of the one-state CSTR model at the operating point, $x_s = \bar y = \bar T = 59.30$ [$^\text{o}$C] and $u_s = \bar F = 630$ [mL/min], results in the following discrete state-space matrices,
\begin{align} \label{eq:linearStateSpaceMatrices}
	A &= \begin{bmatrix}
		0.9572
	\end{bmatrix}, & B &= \begin{bmatrix}
		-57.5381	
	\end{bmatrix}.
\end{align}
Additionally, the measurement function, Eq. (\ref{eq:g}), and output function, Eq. (\ref{eq:h}), are linear and we get,
\begin{align}
	C = C_z = \begin{bmatrix}
		1/V
	\end{bmatrix}.
\end{align} 
The linear-discrete variables, $x_k^l$, $u_k^l$, and $y_k^l$, are deviation variables, i.e., $x_k^l = x(t_k) - x_s$, $u_k^l = u(t_k) - u_s$, and $y_k^l = y(t_k) - y_s$.



\subsection{PI controller with anti-windup as linear control law}

We write the PI controller, Eq. (\ref{eq:PI}), in the linear form, Eq. (\ref{eq:linearController}). We include the integral state of the PI controller as a state in the discrete state-space model. As such, we define,
\begin{align} \label{eq:tilde_x_u}
	\tilde x_k &= \begin{bmatrix}
		x_k^l \\
		I_{k-1}
	\end{bmatrix}, & \tilde u_k &= u_k^l. 
\end{align}
The corresponding state-space matrices are,
\begin{align}
	\hat A &= \begin{bmatrix}
		A & 0 \\
		-T_s k_I C & 1
	\end{bmatrix}, & \hat B &= \begin{bmatrix}
	B \\
	0
	\end{bmatrix}.
\end{align}
The proportional and integral part of the PI controller is linearly expressed as,
\begin{subequations}
	\begin{align}
		\hat K_P &= \begin{bmatrix}
			k_P C & 0 
		\end{bmatrix}, &
		\hat K_I &= \begin{bmatrix}
			T_s k_I C & -1
		\end{bmatrix},
	\end{align}
\end{subequations}
where we point out that the reference is $\bar y^l = \bar y - y_s = 0$. The linear control law is a linear combination of the proportional and integral parts,
\begin{align}
	\hat K &= \hat K_P + \hat K_I.
\end{align}
We express the anti-windup mechanism, Eq. (\ref{eq:PI_aw}), in terms of additional discrete state-space matrices for the integral state,
\begin{align}
	A_{aw} &= \begin{bmatrix}
		0 \\
		T_s k_{aw} \hat K
	\end{bmatrix}, & B_{aw} &= \begin{bmatrix}
	0 \\
	T_s k_{aw}
\end{bmatrix},
\end{align}
such that the state-space matrices for $\tilde x_k$ and $\tilde u_k$ are,
\begin{subequations}
	\begin{align}
		\tilde A &= \hat A + A_{aw}, &
		\tilde B &= \hat B + B_{aw}.
	\end{align}
\end{subequations}
The discrete state-space model is,
\begin{align}
	\tilde x_{k+1} = \tilde A \tilde x_k + \tilde B \tilde u_k,
\end{align}
and the PI response, Eq. (\ref{eq:PI}), is expressed as the linear control response,
\begin{align}
	\tilde u_k = -\hat K \tilde x.
\end{align}

\section{ Monte Carlo based Tuning }
\label{sec:tune}

We apply a MC simulation based method to tune the PI controller, Eq. (\ref{eq:PI}), in the stochastic system, Eq. (\ref{eq:SDE}) \citep{wahlgreen:2021a}. 

\subsection{Objectives for tuning}
We consider two tuning objectives,
\begin{subequations} \label{eq:metricTuning}
	\begin{align}
		\Phi_1^{(n)} &= \sum_{k=0}^N|| z_k^{(n)} - \bar z_k ||_{Q_{z}}^2, \label{eq:metricTuning_1} \\
		\Phi_2^{(n)} &= \sum_{k=0}^N|| z_k^{(n)} - \bar z_k ||_{Q_{z}}^2 + \sum_{n=1}^N || \Delta u_k^{(n)} ||_{Q_{\Delta_u}}^2, \label{eq:metricTuning_2}
	\end{align}
\end{subequations}
where $\bar z_k$ is the output target at sampling time $t_k$, $z_k^{(n)}$ is the output of the n'th closed-loop simulation at sampling time $t_k$, $\Delta u_k^{(n)} = u_k^{(n)} - u_{k-1}^{(n)}$ is the change in input from sampling time $t_{k-1}$ to $t_{k}$, $Q_z$ is an output weight matrix, and $Q_{\Delta u}$ is an input rate of change weight matrix.


\section{Results}
\label{sec:results}

This section presents results for PI tuning, MPC design by controller matching, and closed-loop performance for PI controllers and MPCs. We simulate the stochastic closed-loop system with the three-state CSTR model for $t \in [t_0, t_f]$, where $t_0 = 0$ [s], $t_f = 300$ [s], $R_v = 0.1$, and $\sigma_T = 5$. The sampling time of the system is $T_s = 1$ [s]. We solve the SDE, Eq. (\ref{eq:SDE}), between sampling times with an explicit Euler-Maruyama scheme with $N = 10$ intermediate steps. 

We perform the closed-loop simulations on a 6 core Intel(R) Xeon(R) W-2235 CPU with frequency 3.80GHz. 

\subsection{Initial PI controller and MPC matching}

\begin{figure}[tb]
	\centering
	\includegraphics[trim={0cm 3cm 0cm 3.5cm, clip}, width=0.45\textwidth]{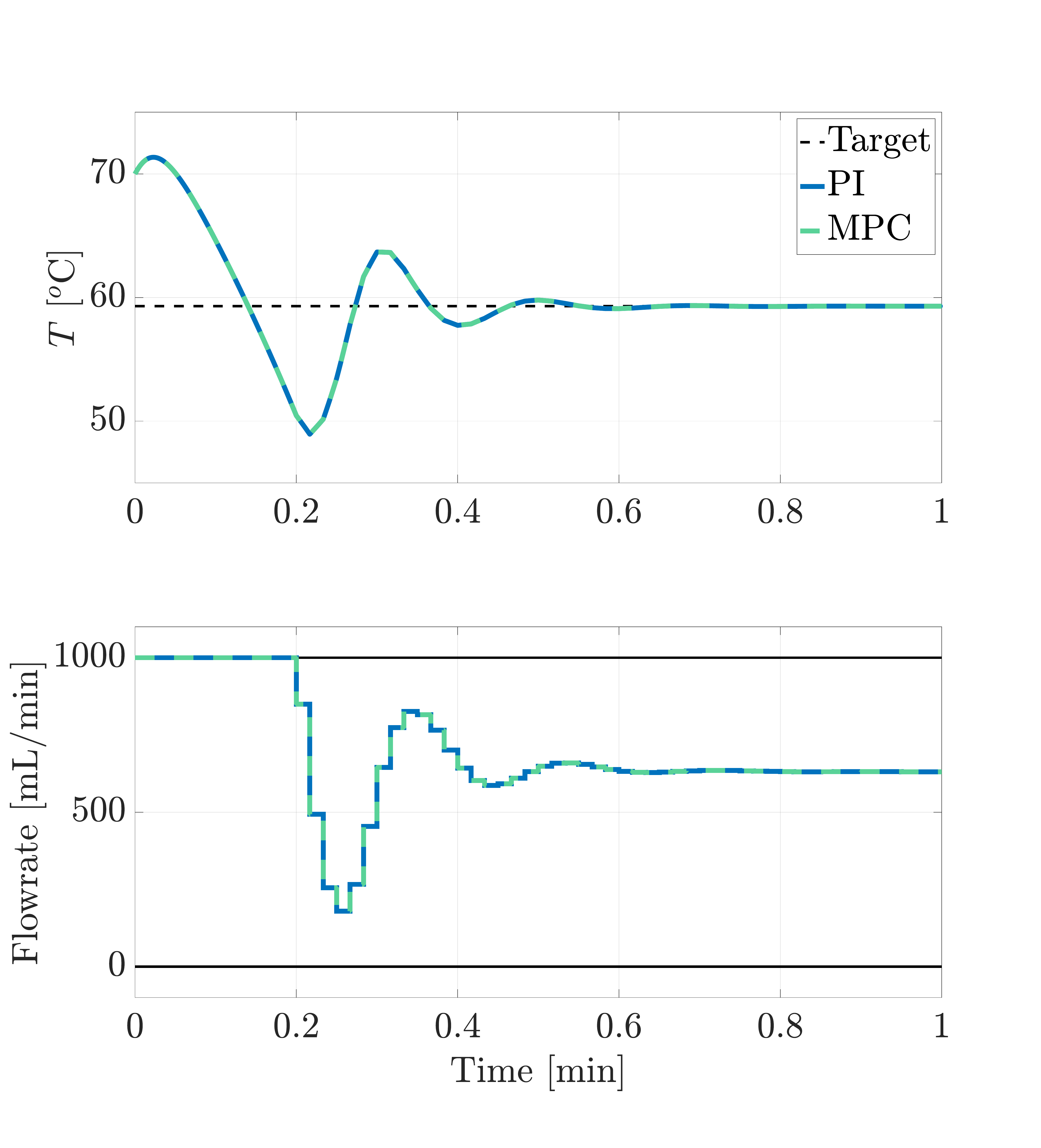}
	\caption{Non-tuned PI controller and matched MPC. The MPC response is identical to the PI response.}
	\label{fig:clopsedLoopDeterministic}
\end{figure}

We initially consider a non-tuned PI controller and demonstrate MPC matching. The PI controller has gains $k_P = -5.0\cdot 10^{-4}$, $k_I = -5.0\cdot 10^{-4}$, and $k_{aw} = 1.0\cdot 10^{-1}$. We obtain the MPC stage cost matrices by solution of the SDP, Eq. (\ref{eq:SDP}). We solve the SDP through cvx with MOSEK in MATLAB \citep{gb08, cvx, mosek}. Figure \ref{fig:clopsedLoopDeterministic} shows that the PI and MPC responses are identical as expected.

\begin{figure}[tb]
	
	\begin{subfigure}{0.45\textwidth}
		\centering
		\includegraphics[trim={0cm 1cm 0cm 1.0cm, clip}, width=\textwidth]{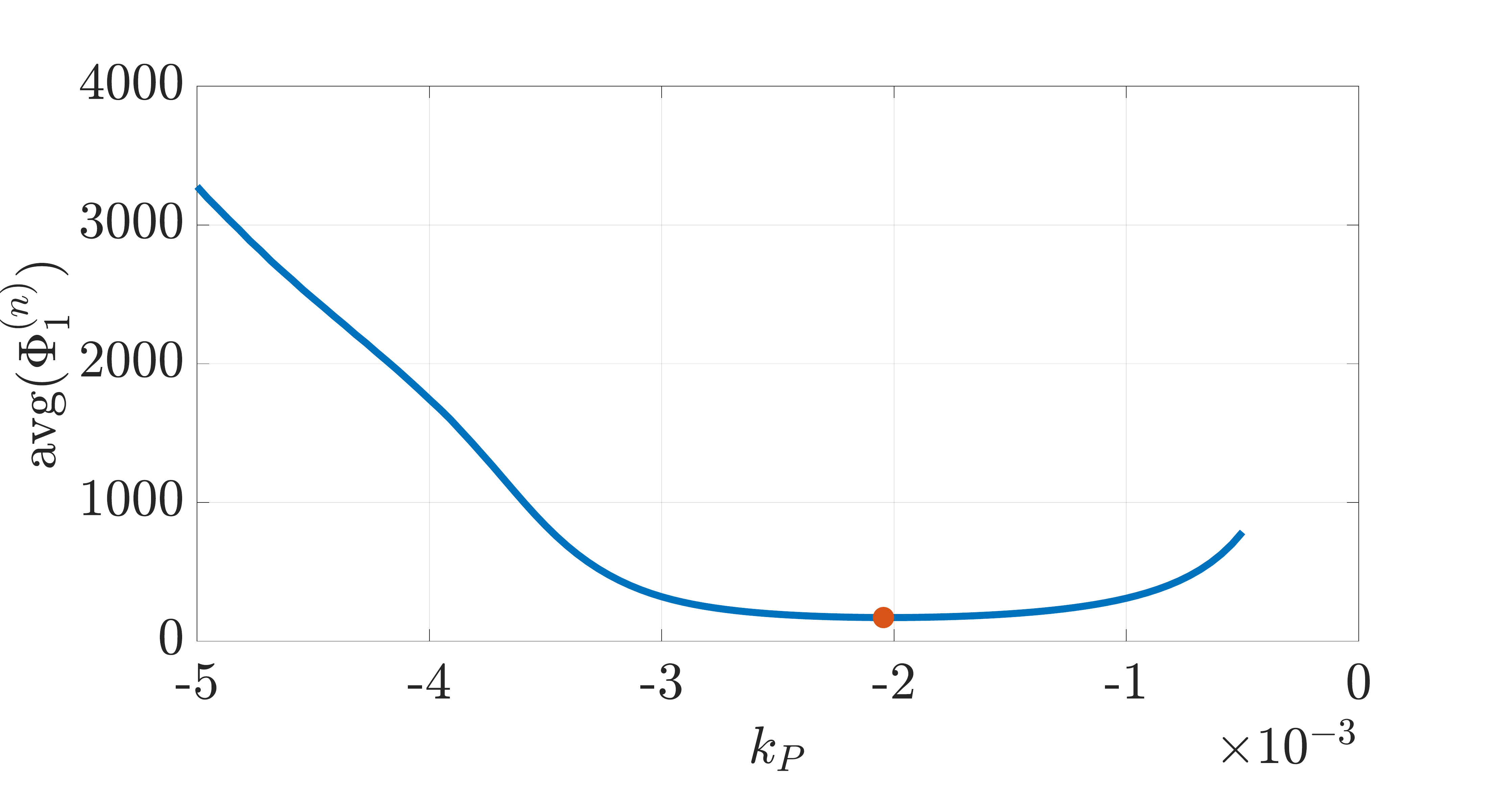}
		\caption{Tuning of $k_P$. The optimum is $k_P = -2.0455\cdot 10^{-3}$. }
		\label{fig:tuneP}
	\end{subfigure}
	\hfill
	\begin{subfigure}{0.45\textwidth}
		\centering
		\includegraphics[trim={0cm 0cm 0cm -1.0cm, clip}, width=\textwidth]{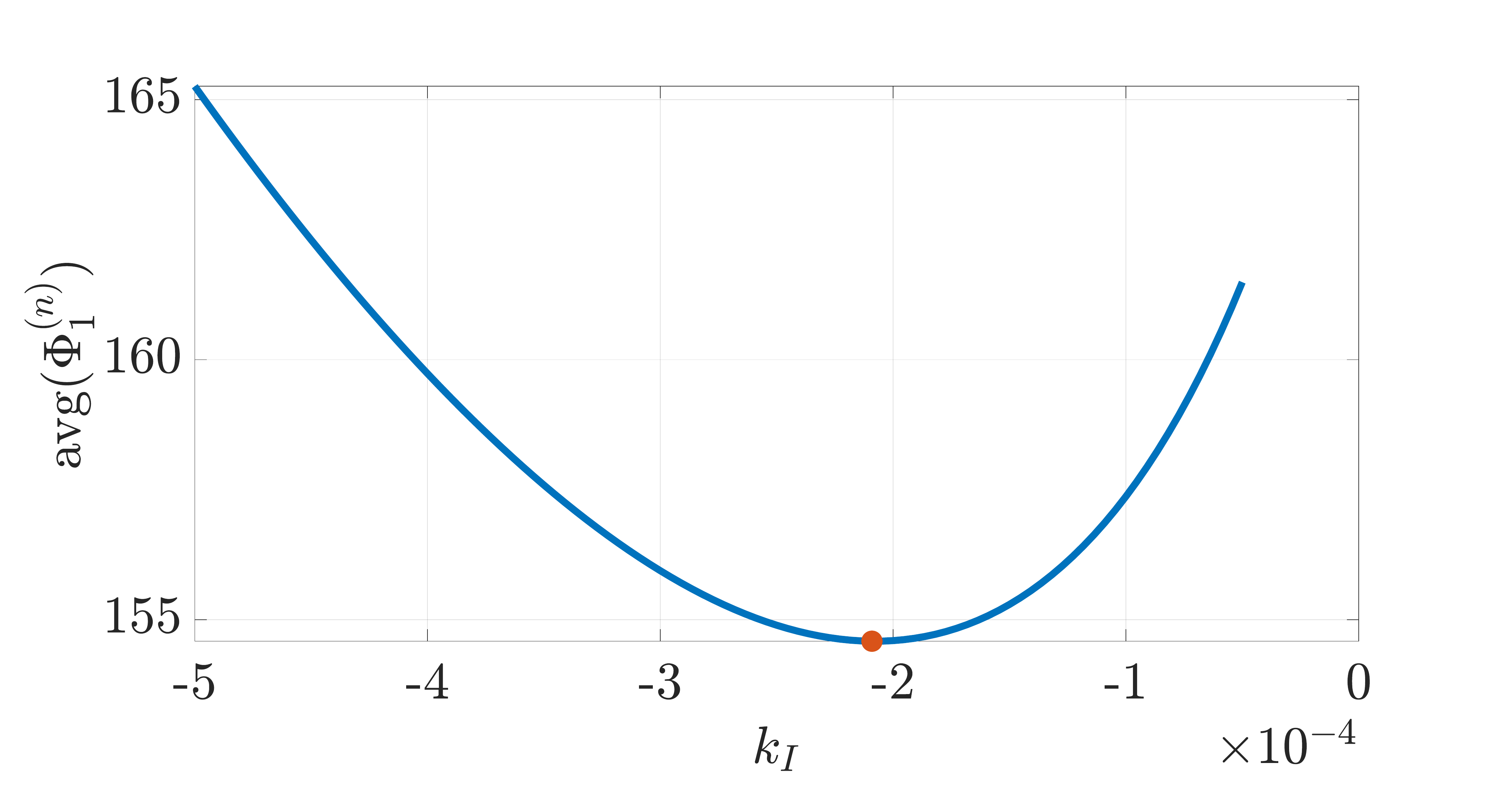}
		\caption{Tuning of $k_I$. The optimum is $k_I = -2.0909\cdot 10^{-4}$. }
		\label{fig:tuneI}
	\end{subfigure}
	\hfill
	\begin{subfigure}{0.45\textwidth}
		\centering
		\includegraphics[trim={0cm 0cm 0cm -1.0cm, clip}, width=\textwidth]{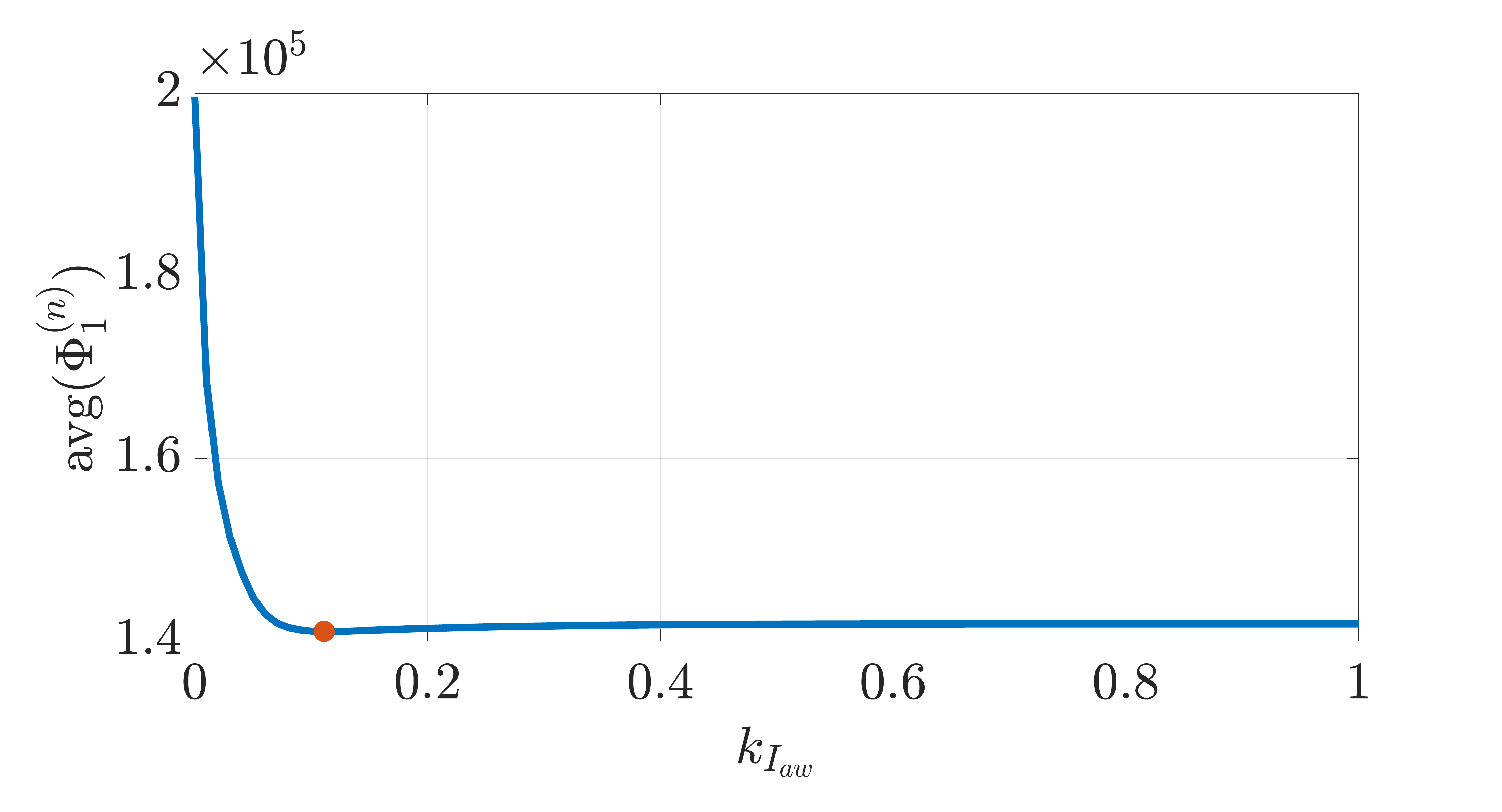}
		\caption{Tuning of $k_{aw}$. The optimum is $k_{aw} = 1.1111\cdot 10^{-1}$. }
		\label{fig:tuneIaw}
	\end{subfigure}
	\caption{ $\Phi_1$ PI gain tuning. Each objective average is computed from $10.000$ closed-loop simulations. }
	\label{fig:tune}
\end{figure}

\begin{figure}[tb!]
	\centering
	\includegraphics[trim={0cm 2cm 0cm 0.0cm, clip}, width=0.45\textwidth]{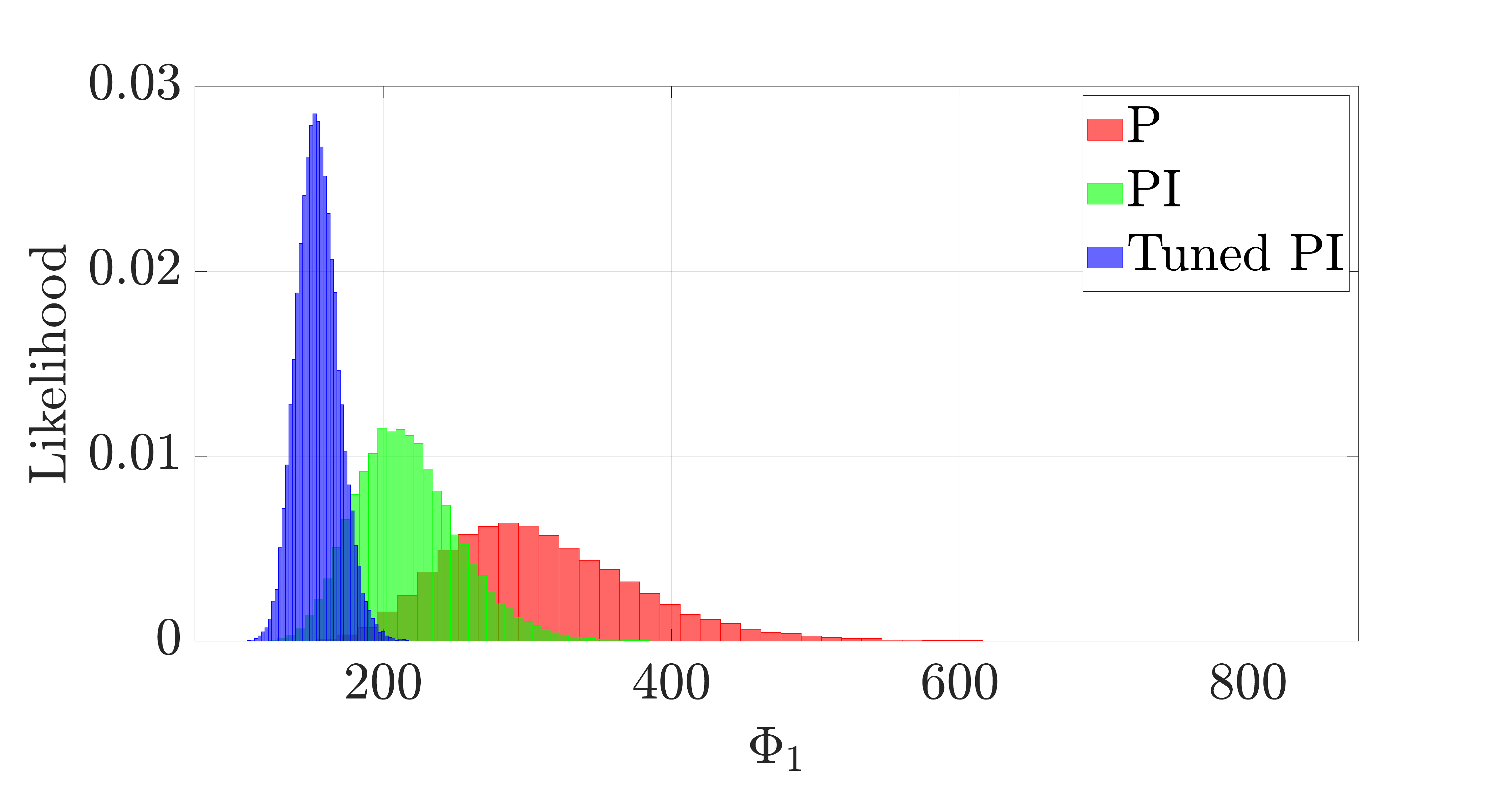}
	\caption{ Probability density function for three controllers. P: $k_P = -1.0\cdot 10^{-3}$. PI: $k_P = -1.0\cdot 10^{-3}$ and $k_I = -1.0\cdot 10^{-4}$. Tuned PI: gains given in Eq. (\ref{eq:PIgains_1}). }
	\label{fig:pdf}
\end{figure}

\begin{figure*}[tb]
	\centering
	\begin{subfigure}{0.48\textwidth}
		\centering
		\includegraphics[trim={0cm 3cm 0cm 3cm, clip}, width=\textwidth]{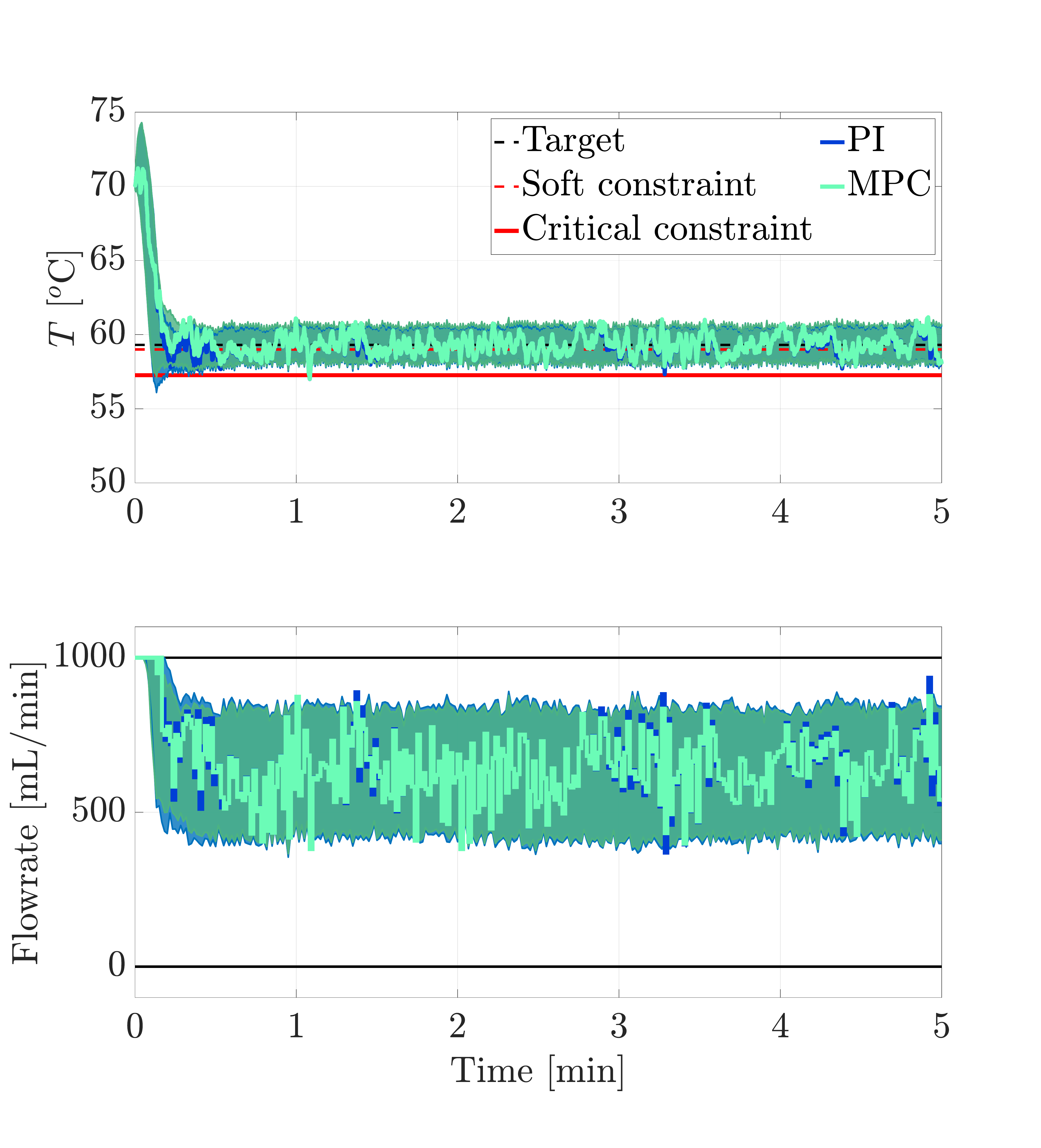}
		\caption{ $\Phi_1$ tuned controllers with chattering input responses. }
		\label{fig:closedLoopTuned}
	\end{subfigure}
	\hfill
	\begin{subfigure}{0.48\textwidth}
		\centering
		\includegraphics[trim={0cm 3cm 0cm 3cm, clip}, width=\textwidth]{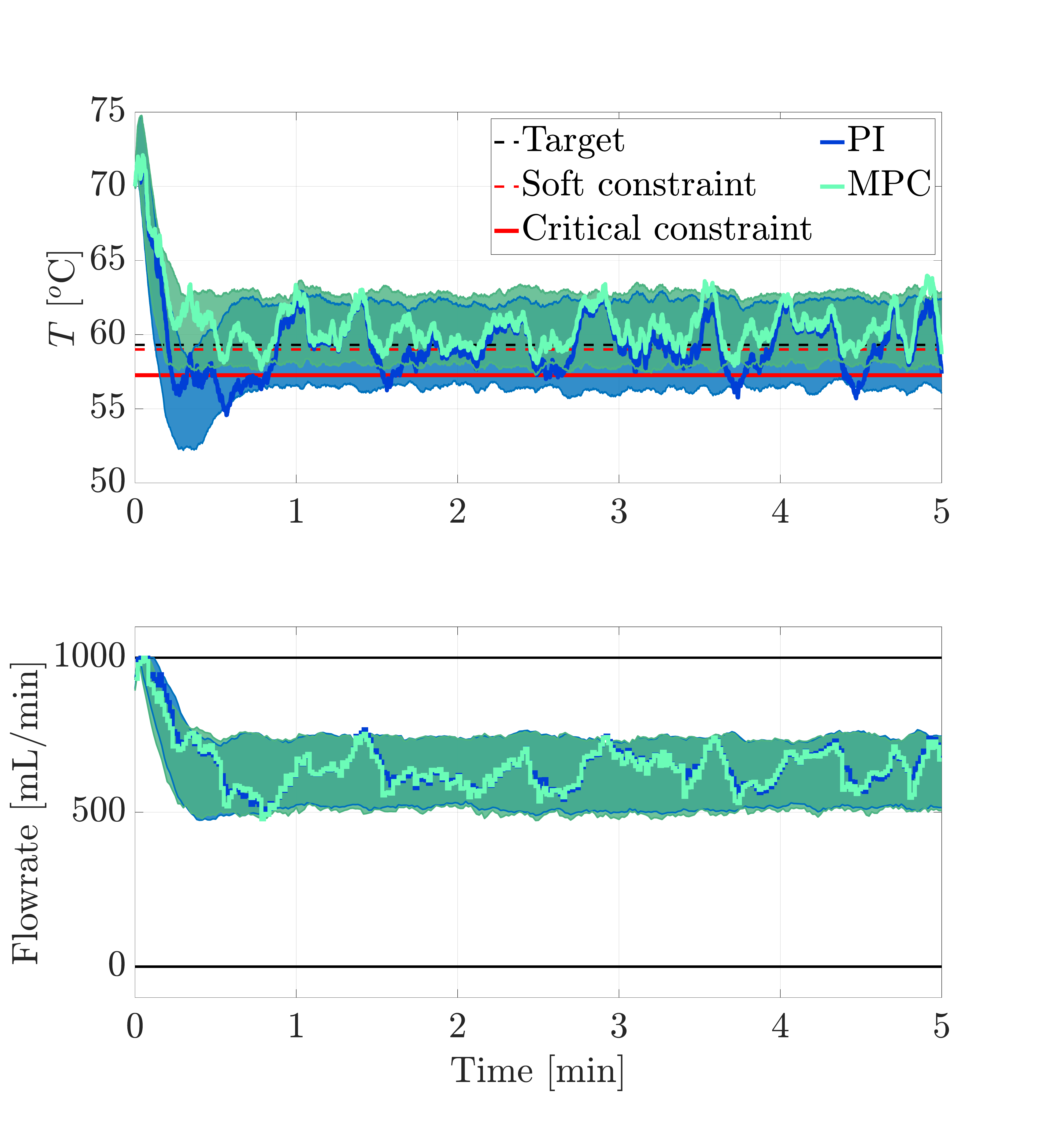}
		\caption{ $\Phi_2$ tuned controllers with less chattering input responses. }
		\label{fig:closedLoopTunedPenalty}
	\end{subfigure}
	\caption{ Highlighted closed-loop trajectories for a single noise realization with PI controllers and matched MPCs and shaded $95\%$ confidence intervals based on 100 simulations. MPC reduces constraint violation compared to the PI controllers. }
	\label{fig:closedLoops}
\end{figure*}


\subsection{Tuning of PI controller}

We base our tuning of each component of the PI gain on $1.000.000$ closed-loop simulation. Thus, tuning of the three PI gains, $k_P$, $k_I$, and $k_{aw}$ requires $3.000.000$ simulations, which we perform in $\approx 90$ [s]. We tune each gain by selecting $100$ equidistant values of the gain in a selected range. The tuned gain value minimizes the average objective value over $10.000$ closed-loop simulations with different process noise. We initialize the system at the operating point for the tuning of $k_P$ and $k_I$, and we initialize the system far from the operating point to tune $k_{aw}$, in order to trigger input saturation. We use $Q_z = 1.0$ and $Q_{\Delta u} = 5\cdot 10^3$.

Figure \ref{fig:tune} presents tuning plots for the gains $k_P$, $k_I$, and $k_{aw}$ with the $\Phi_1$ tuning objective, Eq. (\ref{eq:metricTuning_1}). The tuned PI controller has gains,
\begin{subequations} \label{eq:PIgains_1}
	\begin{align}
		k_P		&= -2.0455\cdot 10^{-3}, &
		k_I		&= -2.0909\cdot 10^{-4}, \\
		k_{aw} 	&=  1.1111\cdot 10^{-1}.
	\end{align}
\end{subequations}
Figure \ref{fig:pdf} presents probability density functions for a P controller, a non-tuned PI controller, and the tuned PI controller based on $30.000$ closed-loop simulations each. We observe that, as expected, the tuned PI controller delivers the best performance both in terms of mean and variance.

We apply the same procedure for the the $\Phi_2$ tuning objective, Eq. (\ref{eq:metricTuning_2}), and obtain the optimal PI gains,
\begin{subequations} \label{eq:PIgains_2}
	\begin{align}
		k_P 	&= -4.0000\cdot 10^{-4}, &
		k_I 	&= -4.9091\cdot 10^{-5}, \\
		k_{aw} 	&=  1.1636\cdot 10^{-1}.
	\end{align}
\end{subequations}


\subsection{Closed-loop simulation with PI controller and MPC}

We consider the MPC formulation, Eq. (\ref{eq:mpc}), with stage costs matched to the tuned PI controllers, Eq. (\ref{eq:PIgains_1}) and Eq. (\ref{eq:PIgains_2}). The input constraints are $u_{\min} = 0$ [mL/min] and $u_{\max} = 1000$ [mL/min]. We impose a lower soft output constraint $z_{\min} = 59.0$ [$^{\text{o}}$C], with $Q_{\eps_l} = 10^6$, and $q_{\eps_l} = 10^6$. Notice that the soft constraint is placed above the critical constraint, $T_{\min}$. This allows the MPC to take action before violation of the critical constraint, while ensuring feasible OCPs. Figure \ref{fig:closedLoops} presents simulation results for both the $\Phi_1$ and $\Phi_2$ tuning objective. The results are based on 100 simulations of the closed-loop system with different noise realizations. The $\Phi_1$ objective leads to large input variance and small output variance compared to the $\Phi_2$ objective. We consider the $\Phi_2$ simulations, where we observe that the increased output variance causes likely constraint violation for the PI controller. The MPC reduces the constraint violation and has an average time out of range of $0.044\%$ compared to the PI controller with $12.18\%$. The $\Phi_1$ values are $1.7079 \cdot 10^3$ and $1.7462 \cdot 10^3$ for the PI controller and MPC respectively, and similarly the $\Phi_2$ values are $2.7160 \cdot 10^3$ and $3.3178 \cdot 10^3$. Thus, the MPC successfully reduces the constraint violation while maintaining the tuned performance of the PI controller at the cost of slightly increased input changes.

\section{Conclusion}
\label{sec:conclusion}

The paper presents a systematic method to design MPC. The method matches the MPC stage cost to a high-performance MC simulation tuned PI controller. As such, the MPC design combines efficient and systematic tuning of a linear controller and advanced MPC properties such as constraint handling. 

We apply the method to design MPC for an exothermic chemical reaction conducted in an adiabatic CSTR, where the operation point is close to a constraint. Our results show, that MPC is successfully matched to the MC simulation tuned PI controller. With the introduction of a soft output constraint, MPC is able to reduce constraint violation compared to the PI controller, while maintaining the tuned performance from the PI controller.


%
%

\bibliographystyle{chicago}
\section{References}


\def\refname{}
\def\bibsection{}

\bibliography{ref/References.bib}%

\begin{thebibliography}{}

\bibitem[\protect\citeauthoryear{Bansal, Sharma, and Shreeraman}{Bansal
  et~al.}{2012}]{bansal:2012}
Bansal, H.~O., R.~Sharma, and P.~R. Shreeraman (2012).
\newblock {PID} {C}ontroller {T}uning {T}echniques: {A} {R}eview.
\newblock {\em Journal of Control Engineering and Technology\/}, 733--764.

\bibitem[\protect\citeauthoryear{Di~Cairano and Bemporad}{Di~Cairano and
  Bemporad}{2009}]{diCairano:2009a}
Di~Cairano, S. and A.~Bemporad (2009, August).
\newblock Model {P}redictive {C}ontroller {M}atching: {C}an {MPC} {E}njoy
  {S}mall {S}ignal {P}roperties of {M}y {F}avorite {L}inear {C}ontroller?
\newblock {\em European Control Conference (ECC)\/}, 2217–2222.

\bibitem[\protect\citeauthoryear{Di~Cairano and Bemporad}{Di~Cairano and
  Bemporad}{2010}]{diCairano:2010a}
Di~Cairano, S. and A.~Bemporad (2010, January).
\newblock Model {P}redictive {C}ontrol {T}uning by {C}ontroller {M}atching.
\newblock {\em IEEE Transactions on Automatic Control\/}~{\em 55\/}(1),
  185–190.

\bibitem[\protect\citeauthoryear{Grant and Boyd}{Grant and Boyd}{2008}]{gb08}
Grant, M. and S.~Boyd (2008).
\newblock Graph implementations for nonsmooth convex programs.
\newblock In V.~Blondel, S.~Boyd, and H.~Kimura (Eds.), {\em Recent Advances in
  Learning and Control}, Lecture Notes in Control and Information Sciences,
  pp.\  95--110. Springer-Verlag Limited.

\bibitem[\protect\citeauthoryear{Grant and Boyd}{Grant and Boyd}{2014}]{cvx}
Grant, M. and S.~Boyd (2014, March).
\newblock {CVX}: Matlab software for disciplined convex programming, version
  2.1.

\bibitem[\protect\citeauthoryear{Jørgensen, Ritschel, Boiroux,
  Schroll-Fleischer, Wahlgreen, Nielsen, Wu, and Huusom}{Jørgensen
  et~al.}{2020}]{jorgensen:2020a}
Jørgensen, J.~B., T.~K.~S. Ritschel, D.~Boiroux, E.~Schroll-Fleischer, M.~R.
  Wahlgreen, M.~K. Nielsen, H.~Wu, and J.~K. Huusom (2020).
\newblock Simulation of {NMPC} for a {L}aboratory {A}diabatic {CSTR} with an
  {E}xothermic {R}eaction.
\newblock {\em Proceedings of 2020 European Control Conference\/}, 202--207.

\bibitem[\protect\citeauthoryear{{MOSEK ApS}}{{MOSEK ApS}}{2022}]{mosek}
{MOSEK ApS} (2022).
\newblock {\em The MOSEK optimization toolbox for MATLAB manual. Version
  9.3.18}.

\bibitem[\protect\citeauthoryear{Qin and Badgwell}{Qin and
  Badgwell}{2003}]{qin:2003}
Qin, S.~J. and T.~A. Badgwell (2003).
\newblock A survey of industrial model predictive control technology.
\newblock {\em Control Engineering Practice\/}~{\em 11}, 733--764.

\bibitem[\protect\citeauthoryear{Wahlgreen, Meyer, Ritschel, Engsig-Karup,
  Gernaey, and Jørgensen}{Wahlgreen et~al.}{2022}]{wahlgreen:2022a}
Wahlgreen, M.~R., K.~Meyer, T.~K.~S. Ritschel, A.~P. Engsig-Karup, K.~V.
  Gernaey, and J.~B. Jørgensen (2022, June 14-17).
\newblock {M}odeling and {S}imulation of {U}pstream and {D}ownstream
  {P}rocesses for {M}onoclonal {A}ntibody {P}roduction.
\newblock {\em The 13th IFAC Symposium on Dynamics and Control of Process
  Systems, including Biosystems (DYCOPS), Busan, Republic of Korea\/}.

\bibitem[\protect\citeauthoryear{Wahlgreen, Reenberg, Nielsen, Rydahl,
  Ritschel, Dammann, and J{\o}rgensen}{Wahlgreen
  et~al.}{2021}]{wahlgreen:2021a}
Wahlgreen, M.~R., A.~T. Reenberg, M.~K. Nielsen, A.~Rydahl, T.~K.~S. Ritschel,
  B.~Dammann, and J.~B. J{\o}rgensen (2021, December 13-17).
\newblock A {H}igh-{P}erformance {M}onte {C}arlo {S}imulation {T}oolbox for
  {U}ncertainty {Q}uantification of {C}losed-loop {S}ystems.
\newblock {\em IEEE Conference on Decision and Control (CDC), Austin, Texas,
  USA\/}.

\bibitem[\protect\citeauthoryear{Wahlgreen, Schroll-Fleischer, Boiroux,
  Ritschel, Wu, Huusom, and Jørgensen}{Wahlgreen
  et~al.}{2020}]{wahlgreen:2020a}
Wahlgreen, M.~R., E.~Schroll-Fleischer, D.~Boiroux, T.~K.~S. Ritschel, H.~Wu,
  J.~K. Huusom, and J.~B. Jørgensen (2020, February 16-19).
\newblock Nonlinear {M}odel {P}redictive {C}ontrol for an {E}xothermic
  {R}eaction in an {A}diabatic {CSTR}.
\newblock {\em 6th Conference on Advances in Control and Optimization of
  Dynamical Systems (ACODS), Chennai, India\/}.

\bibitem[\protect\citeauthoryear{Zanon and Bemporad}{Zanon and
  Bemporad}{2021}]{zanon:2021a}
Zanon, M. and A.~Bemporad (2021).
\newblock {C}onstrained {C}ontrol and {O}bserver {D}esign by {I}nverse
  {O}ptimality.
\newblock {\em IEEE Transactions on Automatic Control, Accepted\/}.

\end{thebibliography}


%
%
%
%
%


\end{document}